\documentclass[12pt]{article}
\usepackage[dvips]{graphicx}
\renewcommand{\baselinestretch}{1.20}
\begin{document}

%
%hep-ph/yymmnnn
\begin{flushright}
OU-HET-657 % , \, August, 2009 \ \ \\
\end{flushright}
\vspace{0mm}
\begin{center}
\large{The role of orbital angular momentum in the proton spin}
\end{center}
\vspace{0mm}
\begin{center}
M.~Wakamatsu\footnote{Email \ : \ wakamatu@phys.sci.osaka-u.ac.jp}
\end{center}
\vspace{-4mm}
\begin{center}
Department of Physics, Faculty of Science, \\
Osaka University, \\
Toyonaka, Osaka 560-0043, JAPAN
\end{center}

%\vspace{4mm}
%PACS numbers : 13.88.+e, 14.20.Dh, 11.10.Hi, 12.39.Ki, 12.38.-t

\vspace{6mm}
\begin{center}
\small{{\bf Abstract}}
\end{center}
\vspace{-2mm}
\begin{center}
\begin{minipage}{15.5cm}
\renewcommand{\baselinestretch}{1.0}
\small
The orbital angular momenta $L^u$ and $L^d$ of up and down
quarks in the proton are estimated as functions of the energy scale
as model-independently as possible, on the basis of Ji's angular
momentum sum rule. This analysis indicates that $L^u - L^d$ is large
and negative even at low energy scale of nonperturbative QCD,
in contrast to Thomas' similar analysis based on the refined
cloudy bag model. We pursuit the origin of this apparent discrepancy
and suggest that it may have a connection with the fundamental
question of how to define quark orbital angular momenta in QCD.

\normalsize
\end{minipage}
\end{center}
\renewcommand{\baselinestretch}{1.2}

\section{Introduction}
\label{intro}

The so-called ``nucleon spin puzzle'' is still one of the most
fundamental problems in hadron physics \cite{EMC88}.
The recent precise measurements of the deuteron spin
structure function by the COMPASS and HERMES groups established that
about $1/3$ of the nucleon spin is carried by the intrinsic quark
polarization \cite{COMPASS07},\cite{HERMES07},
so that the missing spin fraction is now believed to
be of order of $2/3$.
However, there is no widely-accepted consensus on the decomposition
of the remaining part.
(Still, it should be kept in mind that a lot of recent attempts to
directly measure the gluon polarization $\Delta g$ were all led to
the conclusion that $\Delta g$ is likely to be small or at least
it cannot be large enough to resolve the puzzle of the missing nucleon
spin based on the $U_A (1)$ anomaly scenario
\cite{COMPASS06G}\nocite{HERMES07G}\nocite{PHENIX07}-\cite{STAR07}.)

Recently, Thomas claims that the modern spin discrepancy
can well be explained in terms of standard features of the nonperturbative
structure of the nucleon, i.e. relativistic motion of valence quarks,
the pion cloud required by chiral symmetry, and an exchange current
contribution associated with the one-gluon-exchange hyperfine
interaction \cite{Thomas08}\nocite{MT88}\nocite{MT08}-\cite{Thomas09}.
His analysis starts from an estimate of the orbital angular momenta
(OAM) of up and down quarks based on the improved (or fine-tuned)
cloudy bag model taking account of the above-mentioned effects.
Another important factor of his analysis is the observation that
the angular momentum is not a renormalization group invariant quantity,
so that the above predictions of the model should be associated with a
very low energy scale, say, $0.4 \,\mbox{GeV}$.
Then, after solving the QCD evolution equations for the up and down
quark angular momenta, first derived by Ji, Tang and
Hoodbhoy \cite{JTH96}, he was led
to a remarkable conclusion that the orbital angular
momenta of up and down quarks cross over around the scale of
$0.5 \,\mbox{GeV}$. This crossover of $L^u$ and $L^d$ seems absolutely
necessary for his scenario to hold.
Otherwise, the prediction $L^u - L^d > 0$ of the
improved cloudy bag model given at the low energy scale is incompatible
with the current empirical information or lattice QCD simulations at
the high energy scale, which gives $L^u < 0, L^d > 0$.

Actually, the importance of specifying the scale when discussing the
nucleon spin contents, has been repeatedly emphasized in a series of our
papers \cite{WK99}\nocite{WW00}\nocite{Waka03A}\nocite{Waka03B}
\nocite{WN06}-\cite{WN08}. 
(The observation on the scale dependence of the nucleon spin matrix
elements has much longer history. See \cite{Jaffe87} and \cite{JR80},
for instance.)
In particular, we have recently carried out a semi-empirical
analysis of the nucleon spin contents based on Ji's angular momentum
sum rule, and extracted the orbital angular momentum of up and down
quarks as functions of the scale. (See Fig.6 of \cite{WN08}.)
Remarkably, we find no crossover of $L^u$ and $L^d$ when $Q^2$ is varied,
in sharp contrast to Thomas' analysis. This difference is remarkable,
since if there is no crossover of $L^u$ and $L^d$, Thomas' scenario
for resolving the proton spin puzzle is not justified.
The purpose of the present paper is to pursue further the cause of
this discrepancy, which is expected to provide us with a valuable
insight into a very fundamental physical question, i.e. the role of
orbital angular momentum in the nucleon spin.

\section{Semi-empirical extraction of quark orbital angular momenta
in the proton}
\label{sec:1}

There is no doubt about the fact that the nucleon spin consists
of quark and gluon parts as $J^Q + J^g = 1/2$. (Here,
$Q = u + d + s$ for three quark flavors.) The point is that
this decomposition can be made experimentally through the
GPD (generalized parton distribution) analysis of high energy
deeply-virtual Compton scatterings and
of deeply-virtual meson productions. Our semi-phenomenological
estimate of $J^Q$ starts with Ji's angular momentum sum
rule \cite{Ji97},\cite{Ji98} given as 
$J^Q = \frac{1}{2} \,[\,\langle x \rangle^Q + B_{20}^Q (0) \,]$,
where $\langle x \rangle^Q$ is the net momentum fraction carried by
all the quarks, while $B_{20} (0)$ is the net quark contribution to
the anomalous gravitomagnetic moment of the nucleon.
For flavor decomposition, we also need flavor nonsinglet
combinations, i.e.
$J^{(NS)} = \frac{1}{2} \,[\,
 \langle x \rangle^{(NS)} + B_{20}^{(NS)} (0) \,]$,
with $J^{(NS)} = J^{u-d}$, or $J^{u+d-2s}$ etc. 
The quark momentum fractions and the angular momentum fractions are
both scale dependent quantities. Ji showed that they obey
exactly the same evolution equations. At the leading order (LO), the
solutions of the flavor singlet part $J^Q$ is given by
\begin{eqnarray}
 2 \,J^Q (Q^2) &=& \frac{3 \,n_f}{16 + 3 \,n_f} \ + \ \left( 
 \frac{\alpha (Q^2)}{\alpha (Q_0^2)} 
 \right)^{2 \,(16 + 3 \,n_f) / 9 \beta_0} \nonumber \\
 &\,& \hspace{10mm} \times \ 
 \left(\,2 \,J^Q (Q_0^2) - \frac{3 \,n_f}{16 + 3 \,n_f} \,\right),
\end{eqnarray}
with $\beta_0 = 11 - \frac{2}{3} \,n_f$ and similarly for
$\langle x \rangle^Q$. On the other hand,
the scale dependence of the flavor nonsinglet combinations
is given by
\begin{equation}
 2 \,J^{(NS)} (Q^2) \ = \ \left( 
 \frac{\alpha (Q^2)}{\alpha (Q_0^2)} 
 \right)^{32 / 9 \beta_0} \,2 \,J^{(NS)} (Q_0^2),
\end{equation}
and similarly for $\langle x \rangle^{(NS)}$.
%(The orbital angular momenta appearing in the evolution equation
%of Ji, Tang and Hoodbhoy is not gauge invariant quantities.
%However, since the longitudinal quark polarizations are all
%scale invariant at the leading-order level in both of the
%gauge invariant and the chiral invariant factorization scheme,
%the above evolution equation of Ji is practically
%equivalent to that of Ji, Tang and Hoodbhoy at this level.) 

A key observation now is that the quark and gluon momentum fractions are
basically known quantities at least above $Q^2 \simeq 1 \,\mbox{GeV}^2$,
where the framework of perturbative QCD is safely applicable.
For instance, the familiar MRST2004 and CTEQ5 fits give almost the
same quark and gluon momentum fractions below
$10 \,\mbox{GeV}^2$ \cite{MRST},\cite{CTEQ}.
In the following analysis, we shall use the values corresponding to the
scale $Q^2 = 4 \,\mbox{GeV}^2$ from MRST2004 fits.

Neglecting small contribution of strange quarks, which is not essential
for the present qualitative discussion, we are then left with two
unknowns, i.e. $B_{20}^{u+d} (0)$ and $B_{20}^{u-d} (0)$.
For these quantities, we need some theoretical information,
for example, from lattice QCD simulations.
(One must remember the fact that the lattice QCD simulations
at the present stage have a lot of problems, for instance, the omission
of disconnected diagrams, the estimate of the finite volume
effects, and the difficulty of simulations in the realistic chiral
region. The problem is then to judge to what extent we can trust
the predictions of the lattice QCD at the
present stage. This point will be discussed later.)
Fortunately, the available predictions of lattice QCD corresponds
to the renormalization scale $Q^2 \simeq 4 \,\mbox{GeV}^2$, which is high
enough for the framework of perturbative QCD to work. Then, assuming
that all the necessary quantities for the decomposition of the proton
spin are prepared at this high energy scale, an interesting idea is
to use the QCD evolution equations to estimate the corresponding
values at lower energy scales. This is just the
opposite to what was done in Thomas' analysis
\cite{Thomas08},\cite{Thomas09}
as well as in our previous analyses \cite{WN06},\cite{WN08}.
As already mentioned, Thomas uses the
predictions of the improved cloudy bag model as initial
values given at the low energy scale, i.e.
$\sqrt{Q^2} = 0.4 \,\mbox{GeV}$.
Strictly speaking, there is no rigorous theoretical basis for this choice
of starting energy.
It is basically motivated by the fact that a similar scale is needed
to match parton distribution functions calculated in various modern
quark models to high energy experimental data.
An advantage of starting from high energy scale and using
downward evolution is that we can avoid this problem, although the 
precise matching energy with the low energy models are left undetermined. 
Keeping this in mind, one may continue the downward evolution
to the scale $\mu^2$, where $\langle x \rangle^Q = 1$, and
$\langle x \rangle^g = 0$. 
(Numerically, we find that $\mu^2 \simeq 0.070 \,\mbox{GeV}^2$ in the
case the leading-order evolution equation is used, while
$\mu^2 \simeq 0.195 \,\mbox{GeV}$ if the next-to-leading order
evolution equation is used.)
This scale may be regarded as a matching
scale with the low energy effective quark models as advocated in
\cite{MP95} and \cite{Arriola98}. 
Or, one may take a little more conservative viewpoint that the matching
scale would be between $\mu^2$ and somewhere below $1 \,\mbox{GeV}^2$.
At any rate, it is at least obvious that the use of the evolution
equation below this scale, i.e. the unitarity violating limit,
is meaningless.

Now we concentrate on getting reliable information for two unknowns,
i.e. $B_{20}^{u+d} (0)$ and $B_{20}^{u-d} (0)$.
The situation is better for the isovector quantity $B_{20}^{u-d}(0)$.
One finds that the newest predictions of two lattice QCD groups
given at the scale $Q^2 = 4 \,\mbox{GeV}^2$, i.e. 
$B_{20}^{u-d} (0) = 0.274 \pm 0.037$ from the LHPC Collaboration
\cite{LHPC08} and $B_{20}^{u-d} (0) = 0.269 \pm 0.020$ from the
QCDSF-UKQCD Collaboration \cite{QCDSF-UKQCD07},
are remarkably close to each other.
There also exists an estimate based on the chiral quark
soliton model (CQSM). Its prediction evolved to the scale
$Q^2 = 4 \,\mbox{GeV}^2$ from the starting energy scale
$\mu^2 = 0.30 \,\mbox{GeV}^2$ with the next-to-leading (NLO) evolution
equation gives $B_{20}^{u-d} (0) \simeq 0.289$ \cite{WN08}, which
is also close to the lattice QCD estimates of two groups.
To avoid initial scale dependence of the CQSM estimate, we simply
use here the central value of the LHPC Collaboration,
$B_{20}^{u-d} (0) = 0.274$ given at the scale
$Q^2 = 4 \,\mbox{GeV}^2$.

\begin{figure}[htb]
\begin{center}
\includegraphics[height=.35\textheight]{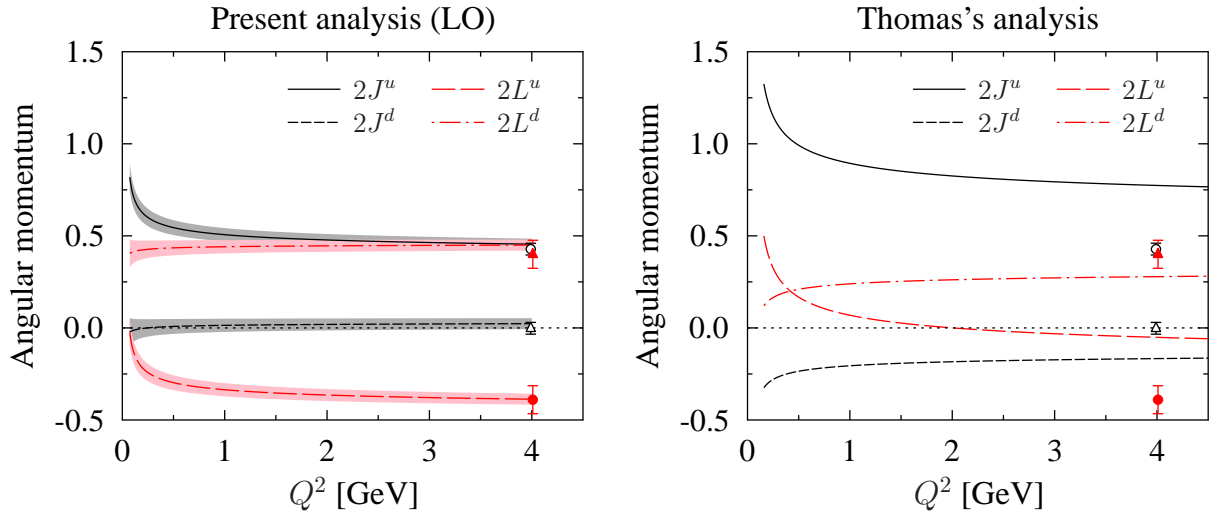}
\caption{The left panel shows the results of the present
semi-phenomenological extraction of the total angular momenta
as well as the orbital angular momenta of up and down quarks,
while the right panel shows the corresponding results of
Thomas \cite{Thomas09}.
In both panels, the open circle, open triangle, filled circle,
and filled triangle respectively
represent the predictions of the LHPC lattice simulations for
$2 \,J^u$, $2 \,J^d$, $2 \,L^u$, and $2 \,L^d$ \cite{LHPC08}.}
\label{Fig:AM}       % Give a unique label
\end{center}
\end{figure}

In contrast to the isovector case, the situation for the isoscalar
combination $B_{20}^{u+d} (0)$ is not very satisfactory, because the
predictions of the lattice QCD simulations are quite sensitive to
the adopted method of chiral extrapolation and dispersed.
The result of the LHPC group obtained with covariant baryon chiral
perturbation theory is $B_{20}^{u+d} (0) = - \,0.094 \pm 0.050$,
while the result of the same group obtained with heavy baryon chiral
perturbation theory is $B_{20}^{u+d} (0) = 0.050 \pm 0.049$.
On the other hand, the result of the QCDSF-UKQCD group is
$B_{20}^{u+d} (0) = - \,0.120 \pm 0.023$.   
Fortunately, from an analysis of the forward limit of the unpolarized
generalized parton distribution $E^{u+d} (x,\xi,t)$ within the
CQSM, the 2nd moment of which gives $B_{20}^{u+d} (0)$,
we were able to give a
reasonable theoretical bound for this quantity, i.e.
$0 \geq B_{20}^{u+d} (0) \geq - \,0.12 \ (\,= \kappa^{p+n}\,)$
with $\kappa^{p+n}$ being the isoscalar magnetic moment of the
nucleon \cite{WN08},
which works to exclude some range of lattice QCD predictions.
In the following, we therefore regard $B_{20}^{u+d} (0)$ as an
unknown constant within this bound. 
(Note that it is a conservative bound since it is actually
given at the low energy model scale and the magnitude of
$B_{20}^{u+d} (0)$ is a decreasing function of the scale parameter
$Q^2$.)

The information on the quark orbital momenta can be obtained from
$J^u$, $J^d$ and $J^s$ by subtracting the corresponding intrinsics
spin contributions, $\Delta \Sigma^u$, $\Delta \Sigma^d$ and
$\Delta \Sigma^s$. Basically, they are all empirically known
quantities. (Note that, at the leading
order, any of these three are scale independent.) 
Among the three combinations $\Delta \Sigma^Q$, $\Delta \Sigma^{u-d}$,
and $\Delta \Sigma^{u+d-2s}$, the flavor singlet one
has a largest uncertainty. For simplicity, here we use the central
value of the recent HERMES analysis, i.e. $\Delta \Sigma^Q = 0.33$,
by neglecting the error-bars.

For completeness, we list below all the initial conditions at
$Q^2 = 4 \,\mbox{GeV}^2$, which we shall use in the present analysis :
\begin{eqnarray}
 \langle x \rangle^Q \!&=&\! 0.579, \ \ 
 \langle x \rangle^{u-d} = 0.158, \ \ 
 \langle x \rangle^s = 0.041, \\
 B_{20}^{u-d} \!&=&\! 0.274, \ \ 
 0 \geq B_{20}^Q = B_{20}^{u+d-2s} 
 \geq - \,0.12, \\
 \Delta \Sigma^Q \!&=&\! 0.33, \ \ 
 \Delta \Sigma^{u-d} = 1.27, \ \ 
 \Delta \Sigma^{u+d-2s} = 0.586 . \ \ 
\end{eqnarray}
(The inclusion of the strange quark contributions to the momentum
fractions and the longitudinal quark polarization appears
inconsistent with the neglect of the corresponding contribution
to $B_{20}$. It is however clear that the influence of the
strange quark components are so small that they
never affect the main point of the present analysis.)

After preparing all the necessary information,
we now evaluate the total angular momentum
as well as the orbital angular momentum of any quark flavor as functions
of $Q^2$. The answers for $2 \,J^u$, $2 \,J^d$ as well
as for $2 \,L^u$, $2 \,L^d$ are shown in the left panel of Fig.1,
respectively by the solid, short-dashed,
long-dashed, and dash-dotted curves with shaded areas.
The open circle, open triangle, filled circle,
and filled triangle in the same figure represent the predictions
of the latest LHPC Collaboration for $2 \,J^u$, $2 \,J^d$, $2 \,L^u$,
and $2 \,L^d$. For comparison, the corresponding
predictions of Thomas' analysis \cite{Thomas08} are shown
in the right panel.
%(The results quoted here are from his newer analysis \cite{Thomas09} 
%not from the original one reported in \cite{Thomas08}.)
One immediately notices that the difference between our analysis and
Thomas' one is sizable. The most significant qualitative
difference appear in the orbital angular momenta.
As already mentioned, Thomas' analysis shows that the orbital
angular momenta of up and down quarks cross over around the scale
of $0.5 \,\mbox{GeV}$. In contrast, no crossover of $L^u$ and $L^d$
is observed in our analysis : $L^d$ remains to be larger than $L^u$
down to the scale where the gluon momentum fraction vanishes.
Comparing the two panels, the cause of this difference seems obvious.
Thomas claims that his results are qualitatively consistent with the
empirical information as well as the lattice QCD data at high energy
scale. (We recall that the sign of $L^{u-d}$ at the high energy scale
is constrained by the asymptotic condition
$L^{u-d} (Q^2 \rightarrow \infty) = - \,\frac{1}{2} \,\Delta \Sigma^{u-d}$,
which is a necessary consequence of QCD
evolution \cite{WN08},\cite{Thomas08}.)
However, the discrepancy between his results and the recent lattice
QCD predictions seems to be never small as is clear from
the right panel of Fig.\ref{Fig:AM}.

\begin{figure}[htb]
\begin{center}
\includegraphics[height=.40\textheight]{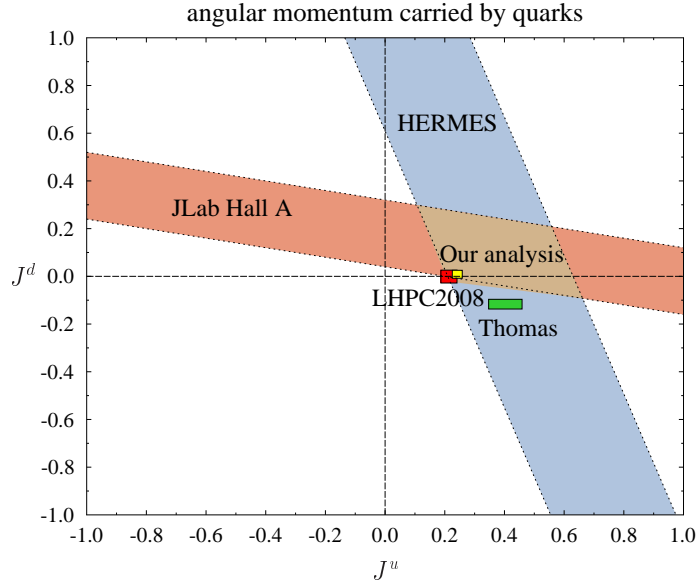}
\caption{The HERMES and JLab Hall A determination of the quark angular
momentum $J^u$ and $J^d$ \cite{HERMES06JA},\cite{HERMES06JB},
\cite{JLab07} in comparison with our semi-empirical
prediction. Also shown for comparison are the recent lattice
QCD prediction by the LHPC Collaboration \cite{LHPC08} and
the result of Thomas' analysis \cite{Thomas09}.}
\label{Fig:JuJd}
\end{center}
\end{figure}

It can also be convinced from a direct comparison with the empirical
information on $J^u$ and $J^d$.
In Fig.\ref{Fig:JuJd}, we compare the prediction
of our semi-empirical analysis, that of Thomas' analysis,
and that of the recent LHPC Collaboration, with the 
HERMES \cite{HERMES06JA},\cite{HERMES06JB} and JLab \cite{JLab07}
determinations of $J^u$ and $J^d$.
One sees that, by construction,
the result of our analysis is fairly close to that of the lattice
QCD simulation.
A slight difference between them comes from the fact
that we use the empirical information (not the lattice QCD
predictions) for the momentum
fractions and the longitudinal polarizations of quarks.
On the other hand, Thomas' result considerably deviates from the
other two predictions.
Although it is consistent with the HERMES data, it lies
outside the error-band of JLab analysis. The latter observation
is mainly related to the fact that his estimate for $J^d$ is
sizably larger than the lattice QCD data or our estimate and
his estimate for $J^d$ is smaller in magnitude than the other two.
(One must be careful about the fact, however, that experimental
extraction of $J^u$ and $J^d$ has a large dependence on the
theoretical assumption of the parametrization of relevant GPDs
and it should be taken as qualitative at the present stage.)

So far, to avoid introducing inessential complexities into
our simple analysis, we did not pay enough care to the
errors of the empirical and semi-empirical information
given at the scale $Q^2 = 4 \,\mbox{GeV}^2$,
except for the quantity $B^{u+d}_{20}(0)$ having the largest uncertainty.
One may worry about how strongly the conclusion of the present analysis
depends on the ambiguities of the other quantities prepared at
$Q^2 = 4 \,\mbox{GeV}^2$. Fortunately, for the isovector quantity
$L^{u-d} \equiv L^u - L^d$, which is of our primary concern in the
present paper, one can convince that our central conclusion
is not altered by these uncertainties.
To see it, let us first recall the relation
\begin{equation}
 2 \,L^{u-d} \ = \ \left[\,\langle x \rangle^{u-d} \ + \ B_{20}^{u-d}(0)
 \,\right] \ - \ \Delta \Sigma^{u-d}.
\end{equation}
Here, $\Delta \Sigma^{u-d} = g_A^{(I=1)}$ is scale independent and
known with high precision, i.e. within $0.3 \,\%$. The momentum
fraction $\langle x \rangle^{u-d}$ is also known with fairly good
precision.
In fact, the difference between the familiar MRST2004 and CTEQ5 fits
at $Q^2 = 4 \,\mbox{GeV}^2$ turns out to be within $1 \,\%$.
The main uncertainty then comes from the isovector anomalous
gravitomagnetic moment of the nucleon $B_{20}^{u-d}(0)$.
We recall again the predictions of the two
lattice QCD collaborations at $Q^2 = 4 \,\mbox{GeV}^2$, i.e.
$B_{20}^{u-d}(0) = 0.274 \pm 0.037$ from the LHPC Collaboration and
$B_{20}^{u-d}(0) = 0.269 \pm 0.020$ from the QCDSF-UKQCD Collaboration,
and also the prediction of the CQSM evolved to the same energy
scale $B_{20}^{u-d}(0) \simeq 0.289$.
In the analysis so far, we have used the central value of the
LHPC prediction by simply neglecting the error-bar. Now let us take
account of the error-bar and see how large this uncertainty would
propagate and affect the value of $L^{u-d}$ at the low energy
model scale. (Note that, the error estimate of the LHPC group is
most conservative and the prediction of the QCDSF-UKQCD group and
that of the CQSM are contained in the error-band of this
LHPC analysis.)

\begin{figure}[htb]
\begin{center}
\includegraphics[height=.40\textheight]{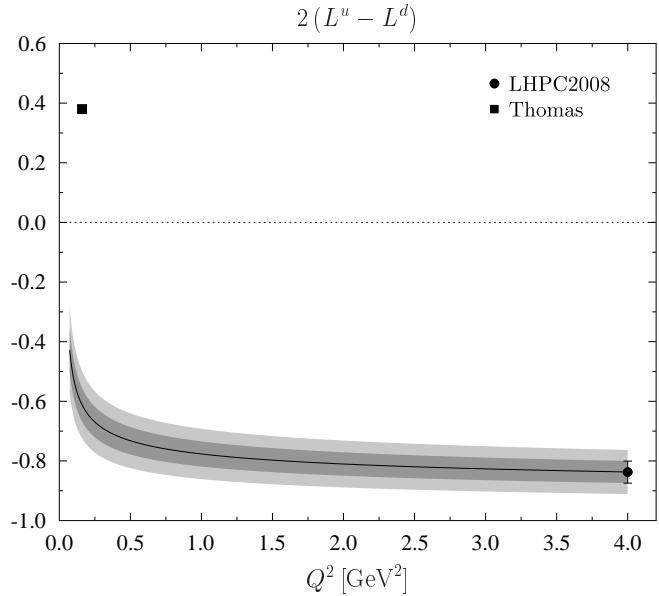}
\caption{The sensitivity of the quark orbital angular momentum
difference $2 \,(L^u - L^d)$ to the initial condition given
at $Q^2 = 4 \,\mbox{GeV}^2$.
The filled area with dark grey is obtained with the LHPC prediction
$B_{20}^{u-d}(0) = 0.274 \pm 0.037$ given at $Q^2 = 4 \,\mbox{GeV}$,
while filled area with light grey is obtained by artificially
doubling the error of LHPC prediction \cite{LHPC08}.
Also shown by the filled square is the prediction of the improved
cloudy bag model corresponding to the scale
$Q_0^2 = 0.16 \,\mbox{GeV}^2$ \cite{Thomas08}.}
\label{Fig:Lu-Ld_err}       % Give a unique label
\end{center}
\end{figure}

The filled area with dark grey in Fig.\ref{Fig:Lu-Ld_err} show the
result of this downward
evolution of $2 \,L^{u-d}$ by starting with the initial condition
given at $Q^2 = 4 \,\mbox{GeV}^2$ on account of this error-band.
In consideration of the possibility of incomplete nature of the present-day
lattice QCD predictions (and also small uncertainties of the other
two quantities $\langle x \rangle^{u-d}$ and $\Delta \Sigma^{u-d}$),
we have also carried out a similar analysis in which the error-bar of
the LHPC prediction is artificially doubled. The result of this
latter analysis is shown by the filled area with light grey.
One can clearly see that the quantity $2 \,L^{u-d}$ remains negative
even down to the lower energy scale close to the unitarity-violating
bound, which appears to be very different from the prediction of the
refined cloudy bag model shown by the filled square in the same figure.

In any case, our semi-phenomenological analysis, which is consistent
with the empirical information as well as the lattice QCD data for
$J^u$ and $J^d$ at high energies, indicates that $L^u - L^d$
remains fairly large and negative even at the low energy scale of
nonperturbative QCD.
If this is in fact confirmed, it may as well be called ``new or
another nucleon spin puzzle''. 
The observation is in fact a serious challenge to any low
energy models of nucleon, since they must now explain small
$\Delta \Sigma^Q$ and large and negative $L^{u-d}$ simultaneously.
The refined cloudy bag model of Thomas and Myhrer appears to be
incompatible with this observation, since it predicts
$2 \,L^u \simeq 0.50$ and $2 \,L^d \simeq 0.12$, or 
$2 \,(L^u - L^d) \simeq 0.38$ at the model scale.
(See TABLE I of \cite{Thomas08}.)
Is there any low energy model which can reproduce this feature ?
Surprisingly, the CQSM can explain
both of these peculiar features of the nucleon observables
at least qualitatively.
It has been long known that it can explain very small
$\Delta \Sigma^Q$ ($\Delta \Sigma^Q \simeq 0.35$ at the model scale)
due to the very nature of the model, i.e. the nucleon as a
rotating hedgehog object \cite{WY91},\cite{BEK88}.
Very interestingly, its prediction for
$L^{u-d}$ given in \cite{WT05}, i.e. $L^{u-d} \simeq - \,0.33$
at the model scale, also matches the conclusion obtained in the
present semi-empirical analysis. (This could be anticipated from
the fact that its prediction for $B_{20}^{u-d}(0)$ matches
the lattice QCD predictions after account of the scale
dependence.)

\section{Note on the nucleon spin decomposition}
\label{sec:3}

To understand the cause of the apparent mismatch between our
observation and the picture of standard quark models, typified by
the refined cloudy bag model,
it may be of some help to remember the important fact that the
decomposition of the nucleon spin is not unique at all.
There are two widely-known decompositions of the nucleon spin,
i.e. the Ji decomposition \cite{Ji97}
and the Jaffe-Manohar one \cite{JM90},\cite{BJ98}.
(See also recent yet another proposal \cite{Chen08},\cite{Chen09}.)
The Ji decomposition is given in the form
\begin{equation}
 \frac{1}{2} \ = \ J^Q \ + \ J^g ,
\end{equation}
whose terms are defined as nucleon matrix elements of the corresponding
operators
\begin{eqnarray}
 \,
 \hat{\mbox{\boldmath $J$}}^Q &=& \int \,\psi^\dagger \,\left[\,
 \frac{1}{2} \,\,\mbox{\boldmath $\Sigma$}
 \ + \ \mbox{\boldmath $x$} 
 \times (\, - \,i \,\mbox{\boldmath $D$} \,)
 \,\right] \,\psi \,d^3 x , \\
 \hat{\mbox{\boldmath $J$}}^g &=& \int \,
 \left[\, \mbox{\boldmath $x$} \times 
 (\,\mbox{\boldmath $E$} \times \mbox{\boldmath $B$} \,) \,\right]
 \,d^3 x ,
\end{eqnarray}
with $\mbox{\boldmath $\Sigma$} = \gamma^0 \,\mbox{\boldmath $\gamma$} \,
\gamma^5$. On the other hand, the Jaffe-Manohar decomposition is given
in the form
\begin{equation}
 \frac{1}{2} \ = \ J^{\prime Q} \ + \ J^{\prime g} ,
\end{equation}
whose terms are defined as nucleon matrix elements of the following
operators
\begin{eqnarray}
 \,
 \hat{\mbox{\boldmath $J$}}^{\prime Q} &=& \int \,
 \psi^\dagger \,\left[\,\,\frac{1}{2} \,\,
 \mbox{\boldmath $\Sigma$} \ + \ \mbox{\boldmath $x$} 
 \times (\, - \,i \,\nabla \,)
 \,\right] \,\psi \,d^3 x , \\
 \hat{\mbox{\boldmath $J$}}^{\prime g} &=& \int \,
 \left[\, (\,\mbox{\boldmath $E$} \times \mbox{\boldmath $A$} \,) \ - \ 
 E_i \,(\mbox{\boldmath $x$} \times \mbox{\boldmath $E$} \,) \,A_i
 \,\right] \,d^3 x .
\end{eqnarray}

Very recently, Burkardt and BC have studied both the Jaffe-Manohar
as well as the Ji decomposition of angular momentum for two simple toy
models, i.e. in the scalar diquark model as well as for an electron
in QED in order $\alpha = e^2 / 4 \,\pi$ \cite{BBC08}.
They demonstrated that both decomposition yield the same numerical value
for the fermion OAM in the scalar diquark model, but not in QED.
They have also shown that the fermion OAM
distributions in the Feynman-$x$ space in both decompositions dot not
coincide even in the scalar diquark model.

Their investigation throws a renewed interest in the difference existing
between the quark OAM resulting from the Jaffe-Manohar decomposition
and that obtained from the Ji decomposition.
It has been long recognized that the quark OAM in the Ji decomposition
is manifestly gauge invariant, and accordingly it contains an interaction
term with the gluon. On the other hand,
the quark OAM appearing in the Jaffe-Manohar decomposition has
simpler physical interpretation as a canonical orbital angular
momentum in that it is given as a nucleon matrix element of free-field
expression of quark OAM. Unfortunately, no reliable information exists
on the difference between the magnitudes of these two definitions of
the quark OAMs from lattice QCD simulations.

Since the CQSM is an effective quark theory that contains no gauge
field, one might naively expect that there is no such ambiguity
problem in the definition of the quark OAM.
It turns out that this is not necessarily the case, however.
The point is that it is a highly nontrivial interaction
theory of quark fields. To explain it, we recall the past analyses
of Ji's angular momentum sum rule within the framework of the CQSM.
The analysis for the isoscalar combination was carried out by
Ossmann et al. \cite{Ossmann05}.
Starting with the theoretical expression for the
unpolarized GPD $E_M^{u+d} (x,\xi,t) \equiv H^{u+d} (x,\xi,t) + 
E^{u+d} (x,\xi,t)$, they analyzed its 2nd moment, which is expected
to give $2 \,J^{u+d}$ on the basis of general argument of Ji.
In fact, by using the equation of motion of the model, they could show
that
\begin{eqnarray}
 \frac{1}{2} \,\int_{- 1}^1 \,x \,E^{u+d} (x,0,0) \,dx \ = \ 
 L_f^{u+d} \ + \ \frac{1}{2} \,\Delta \,\Sigma^{u+d} ,
\end{eqnarray}
where the terms in the r.h.s are respectively given as proton
(with spin up) matrix elements of the following operators within
the model : 
\begin{eqnarray}
 \hat{L}_f^{u+d} &=& \int \,\psi^\dagger (x) \,\left[\,
 \mbox{\boldmath $x$} \times (- \,i \,\nabla) \,\right]_3
 \,\psi (x) \,d^3 x , \\
 \hat{\Sigma}^{u+d} &=& \int \,\psi^\dagger (x) \,\Sigma_3 \,
 \psi (x) \,d^3 x .
\end{eqnarray}
As anticipated, the answer is given as a sum of the proton matrix
element of the free-field expression for the quark OAM operator and
that of the isoscalar quark spin operator. This is nice, but still
we must be careful about the following fact. The net quark OAM
distribution in $x$-space defined through the unintegrated version of
Ji's sum rule written as
\begin{equation}
 L^{u+d} (x) \ = \ \frac{1}{2} \,x \,E_M^{u+d} (x,0,0) \ - \ 
 \frac{1}{2} \,\Delta \Sigma^{u+d} (x) ,
\end{equation}
does not seem to coincide with the OAM distribution corresponding to
the Jaffe-Manohar decomposition numerically evaluated within the same
CQSM in \cite{WW00}. This observation just corresponds to the recent
finding by Burkardt and BC in the scalar diquark model \cite{BBC08}.

A similar analysis for the isovector combination was carried out
in \cite{WT05}.
It was found there that the 2nd moment of the isovector
GPD $E_M^{u-d} (x,0,0)$
is now given as a sum of three pieces as
\begin{eqnarray}
 \frac{1}{2} \,\int_{-1}^1 \,x \,E_M^{u-d} (x,0,0) \,dx &=& 
 \left(\, L_f^{u-d} \ + \ \delta L^{u-d} \,\right) \nonumber \\
 &\,& + \ \ \frac{1}{2} \,\Delta \Sigma^{u-d} .
\end{eqnarray}
Here, $L_f^{u-d}$ and $\Delta \Sigma^{u-d}$ terms are naively anticipated
ones, i.e. a proton matrix element of free-field expression for the
isovector quark OAM operator and that of the isovector quark spin
operator respectively given as
\begin{eqnarray}
 \hat{L}_f^{u-d} &=& \int \,\psi^\dagger (x) \,\tau_3 \,
 \left[\, \mbox{\boldmath $x$} \times (- \,i \,\nabla ) \,\right]_3 \,
 \psi (x) \,d^3 x , \\
 \hat{\Sigma}^{u-d} &=& \int \,\psi^\dagger (x) \,\tau_3 \,
 \Sigma_3 \,\psi (x) \,d^3 x .
\end{eqnarray}
Somewhat embarrassingly, we found an extra piece
represented as
\begin{eqnarray}
 &\,& \delta L^{u-d} \ = \ - \,M \,\,\frac{N_c}{18} \nonumber \\
 &\,& \times \sum_{n \in occ} \,
 \langle n \,| \, r \,\sin F(r) \,
 \gamma^0 \,[\,\mbox{\boldmath $\Sigma$} \cdot 
 \hat{\mbox{\boldmath $r$}}
 \,\mbox{\boldmath $\tau$} \cdot \hat{\mbox{\boldmath $r$}} - 
 \mbox{\boldmath $\Sigma$} \cdot \mbox{\boldmath $\tau$} \,
 \,] \,| \,n \rangle . \ \ \ 
\end{eqnarray}
Here, $| n \rangle$ stand for the eigenstates of the Dirac Hamiltonian
$H = - \,i \,\mbox{\boldmath $\alpha$} \cdot \nabla + M \,\beta \,
e^{\,i \,\gamma_5 \,\mbox{\boldmath $\tau$} \cdot \hat{\mbox{\boldmath $r$}}
\,F(r)}$ with hedgehog mean field.
The symbol $\sum_{n \in occ}$
denotes the sum over all the occupied eigenstates of $H$.
This extra term is highly model-dependent and its physical interpretation
is far from self-evident.
It is nevertheless clear that there is no compelling reason to
believe that the quark OAM defined through Ji's sum rule
must coincide with the canonical one, i.e. the proton matrix
element of the free-field OAM operator. Since the CQSM is a nontrivial
interacting theory of effective quarks, which mimics the important
chiral-dynamics of QCD, it seems natural to interpret this peculiar
term as a counterpart of the interaction dependent part of the quark
OAM in the Ji decomposition of the nucleon spin.

%\vspace{3mm}
%\newcommand{\lw}[1]{\smash{\lower2.ex\hbox{#1}}}
%\begin{table}[h]
%\caption{The CQSM predictions for $L_f^{u-d}$ and $\delta L^{u-d}$
%as well as their sum at the leading order in the collective angular velocity
%$\Omega$.}
%\label{Table:quarkOAM}
%\vspace{2mm}
%\begin{center}
%\renewcommand{\arraystretch}{1.0}
%\begin{tabular}{cccc}
%\hline\hline
% & \ \ \ \ $L_f^{u-d}$ \ \ \ & \ \ $\delta L^{u-d}$ \ \ &
% \ \ \ \ \ $L^{u-d} = L_f^{u-d} + \delta L^{u-d}$ \ \ \\
% \hline
% \ \ Valence \ \ \ & \ 0.147 & - \,0.289 & - \,0.142 \\
% \hline
% \ \ Sea \ \ \ & - \,0.265 & \ 0.077 & - \,0.188 \\
% \hline
% \ \ Total \ \ \ & - \,0.115 & - \,0.212 & - \,0.330 \\
%\hline \hline
%\end{tabular}
%\end{center}
%\end{table}

\begin{table}
\begin{center}
\caption{The CQSM predictions for $L_f^{u-d}$ and $\delta L^{u-d}$
as well as their sum at the leading order in the collective angular
velocity $\Omega$.}
\label{Table:quarkOAM}       % Give a unique label
\vspace{8mm}
\begin{tabular}{cccc}
\hline\hline
 & \ $L_f^{u-d}$ \ & \ $\delta L^{u-d}$ \ &
 \ $L^{u-d} = L_f^{u-d} + \delta L^{u-d}$ \ \\
 \hline
 \ Valence \ \ & \ 0.147 & - \,0.289 & - \,0.142 \\
 \hline
 \ Sea \ \ & - \,0.265 & \ 0.077 & - \,0.188 \\
 \hline
 \ Total \ \ & - \,0.115 & - \,0.212 & - \,0.330 \\
\hline \hline
\end{tabular}
% Or use
% \vspace*{5cm}  % with the correct table height
\end{center}
\end{table}

A natural next question is how significant the influence of
this peculiar term is. For illustration, we show in table 1 the predictions
of the CQSM for $L_f^{u-d}$ and $\delta L^{u-d}$ as well as their sum.
(The numerical values are from the leading-order prediction of the CQSM
given in \cite{WT05}.) 
Here, shown in the 2nd and the 3rd rows are respectively
the contributions of the three valence quarks and those of the negative
energy Dirac-sea quarks, while shown in the 4th row are their sums.
One sees that the valence quark contribution to $L_f^{u-d}$ is positive
but the Dirac-sea contribution to it is negative and larger in magnitude
than the valence quark one, so that the net
contribution to $L_f^{u-d}$ is negative. Concerning the term
$\delta L^{u-d}$, it is dominated by the valence quark contribution,
which is large and negative.
Adding up the two contributions, $L_f^{u-d}$ and $\delta L^{u-d}$,
we thus find that the CQSM prediction for the isovector quark OAM
$L^u - L^d$ is sizably negative. We again emphasize that this prediction
of the CQSM is totally different from the corresponding prediction of the
refined CB model, which gives that $L^u - L^d$ is sizably large and
positive at the model scale. A word of caution here.
The calculation of the
quark OAM in the CB model does not seem to use Ji's way of defining the
quark OAM, although the detail is not clear from the papers.
It would be interesting if one can check whether the two ways of
calculating the quark orbital angular momenta make any difference also
in the framework of the refined CB model or not.

\section{Summary and Conclusion}
\label{sec:4}
To sum up, we have estimated the orbital angular momenta $L^u$
and $L^d$ of up and down quarks in the proton as functions
of the energy scale, by carrying out a downward QCD evolution
of available information at high energy, to find that
$L^u - L^d$ remains to be large and negative even at the low
energy scale of nonperturbative QCD, in remarkable contrast to
Thomas' conclusion based on the refined cloudy bag model.
Although the orbital angular momenta of quarks are not direct
observables, they can well be extracted since $J^u$ and $J^d$ are
measurable quantities from GPD analysis and since the intrinsic
quark polarizations are basically known quantities by now.
(One should not forget about the fact that the orbital angular
momenta of quarks extracted in this way correspond to the Ji
decomposition.)
Then, what is required for future experiments is to determine
$J^u$ and $J^d$ as precisely as possible including their
scale dependence. Ideal would be to confirm the
predicted strong scale dependence between $1 \,\mbox{GeV}$ and
several hundreds $\mbox{MeV}$ region. In practice, the GPD
analysis far below $1 \,\mbox{GeV}$ may not be so easy because of
uncontrollable higher-twist effects. However, the precise
determination of $J^u$ and $J^d$ around $1 \,\mbox{GeV}$ region
should give crucial information to judge which of the two scenarios,
Thomas' one and the present one, for $L^u$ and $L^d$, are close to
the truth, thereby providing us with a valuable insight into
unexpected role of quark orbital angular momenta as ingredients
of the nucleon spin. 

\vspace{16mm}
%\newpage
\noindent
\begin{large}
{\bf Acknowledgment}
\end{large}

\vspace{3mm}
This work is supported in part by a Grant-in-Aid for Scientific
Research for Ministry of Education, Culture, Sports, Science
and Technology, Japan (No.~C-A215402680)

%%
%% For one-column wide figures use
%\begin{figure}
%% Use the relevant command for your figure-insertion program
%% to insert the figure file.
%% For example, with the option graphics use
%\resizebox{0.75\textwidth}{!}{%
%  \includegraphics{leer.eps}
%}
%% If not, use
%%\vspace{5cm}       % Give the correct figure height in cm
%\end{figure}
%%

%% For two-column wide figures use
%\begin{figure*}
%% Use the relevant command for your figure-insertion program
%% to insert the figure file. See example above.
%% If not, use
%\vspace*{5cm}       % Give the correct figure height in cm
%\end{figure*}
%%

%\begin{table}
%\begin{tabular}{lll}
%\hline\noalign{\smallskip}
%\end{tabular}
%% Or use
%\vspace*{5cm}  % with the correct table height
%\end{table}

%%
%% BibTeX users please use
%% \bibliographystyle{}
%% \bibliography{}
%%
%% Non-BibTeX users please use
%\begin{thebibliography}{}
%%
%% and use \bibitem to create references.
%Author, \textit{Book title} (Publisher, place year) page numbers
%\end{thebibliography}

\end{document}